\documentclass[lettersize,journal]{IEEEtran}
\usepackage{amsmath,amsfonts}
\usepackage{algorithmic}
\usepackage{algorithm}
\usepackage{array}
\usepackage{array,multirow,graphicx}
\usepackage[caption=false,font=normalsize,labelfont=sf,textfont=sf]{subfig}
\usepackage{textcomp}
\usepackage{stfloats}
\usepackage{url}
\usepackage[dvipsnames]{xcolor}
\usepackage{verbatim}
\usepackage{graphicx}
\usepackage{cite}
\usepackage{caption}
\usepackage{nomencl}
\usepackage{amsmath}
\usepackage{enumitem}
\usepackage{hyperref}
\makenomenclature
\begin{document}
\title{Preventive Energy Management for Distribution Systems Under Uncertain Events: A Deep Reinforcement Learning Approach}

\author{Md Isfakul Anam,~\IEEEmembership{~Clarkson University}, Tuyen Vu,~\IEEEmembership{~Clarkson University}, Jianhua Zhang, ~\IEEEmembership{~Clarkson University}
\thanks{Md Isfakul Anam, and T. Vu with Clarkson University, Potsdam, NY, USA; Emails: isfakum@clarson.edu, tvu@clarkson.edu. Corresponding Author: T. Vu, Email: tvu@clarkson.edu}}

\maketitle

\begin{abstract}
As power systems become more complex with the continuous integration of intelligent distributed energy resources (DERs), new risks and uncertainties arise. Consequently, to enhance system resiliency, it is essential to account for various uncertain events when implementing the optimization problem for the energy management system (EMS). This paper presents a preventive EMS considering the probability of failure (PoF) of each system component across different scenarios. A conditional-value-at-risk (CVaR)-based framework is proposed to integrate the uncertainties of the distribution network. Loads are classified into critical, semi-critical, and non-critical categories to prioritize essential loads during generation resource shortages. A proximal policy optimization (PPO)-based reinforcement learning (RL) agent is used to solve the formulated problem and generate the control decisions. The proposed framework is evaluated on a notional MVDC ship system and a modified IEEE 30-bus system, where the results demonstrate that the PPO agent can successfully optimize the objective function while maintaining the network and operational constraints. For validation, the RL-based method is benchmarked against a traditional optimization approach, further highlighting its effectiveness and robustness. This comparison shows that RL agents can offer more resiliency against future uncertain events compared to the traditional solution methods due to their adaptability and learning capacity.

\end{abstract}


\nomenclature[00]{\( w_i\)}{Weight of load $i$}
\nomenclature[01]{\(P_{i,g}, Q_{i,g}\)}{Real and reactive power output of $i$-th generator}
\nomenclature[02]{\(P_{i,l}, Q_{i,l}\)}{Real and reactive load at $i$-th bus}
\nomenclature[03]{\(P_{i,inj}, Q_{i,inj}\)}{Real and reactive power injection at $i$-th bus}
\nomenclature[04]{\(V_i, \theta_{ik}\)}{voltage magnitude and voltage angle difference at $i$-th bus}
\nomenclature[05]{\(G_{ik}, B_{ik}\)}{Conductance and susceptance of line $ik$}
\nomenclature[06]{\(P_{ik}^{lim}, Q_{ik}^{lim}\)}{Real and reactive power capacity of line $ik$}
\nomenclature[07]{\(P_{ik}, Q_{ik}\)}{Real and reactive power flowing through line $ik$}
\nomenclature[08]{\(P_i^{C,max}, P_i^{D,max}\)}{Maximum charging and discharging rate of $i$-th ESS}
\nomenclature[09]{\(E_{i,b}\)}{Energy stored at $i$-th ESS}
\nomenclature[10]{\(P_{i,b}^r\)}{Output power of $i$-th ESS}
\nomenclature[11]{\(SOC_i\)}{State of charge of $i$-th ESS}
\nomenclature[12]{\(\alpha_s\)}{Probability of scenario $s$}
\nomenclature[13]{\(N_g\)}{Number of generator}
\nomenclature[14]{\(T\)}{Total planning horizon}
\nomenclature[15]{\(\triangle t\)}{Time step}
\nomenclature[16]{\(B\)}{Number of buses}

\printnomenclature

\section{Introduction}


As the energy demand is increasing around the world, the power network is being expanded, and advanced technologies are being introduced in this sector to ensure reliable and uninterrupted power supply. However, with the increasing expansion and complexity, the power system is getting exposed to different uncertain events such as cyber-attacks, faults, and extreme weather conditions. These events can lead to severe damage to the power systems, causing major power outages and component failures \cite{boteler2001space}. As a result, these uncertain events should be considered in power system operations to improve the system's resiliency. 

In power systems, resilience refers to the grid's capacity to anticipate and adjust to shifting operational conditions, as well as to endure and quickly recover from significant disruptions due to natural incidents or intentional cyber-physical attacks \cite{directive2013critical}. According to the National Infrastructure Advisory Council (NIAC) in the USA, resilience also involves the ability to learn from past disturbances and make adjustments to operations or infrastructure to lessen the impact of future similar events \cite{berkeley2010framework}. Grid resilience can be categorized into three phases; pre-event or preventive phase, disruption phase, and post-event or restorative phase \cite{panteli2015grid}. As a result, enhancing grid resilience requires a multi-stage approach that includes preventive measures before extreme events, corrective actions during the disaster, and restorative efforts to recover after disruptions \cite{umunnakwe2021quantitative}. This paper focuses on the preventive actions of an Energy Management System (EMS) considering future uncertain events. 

Pre-disaster actions or preventive measures are crucial for power system resilience because they help the grid prepare for potential disruptions, mitigate the severity of their impact, and reduce the cost and time of restoration \cite{en17010168}. Preventive measures can be classified into two broad categories: grid hardening and operational strategies \cite{7922545}. Hardening measures focus on improving the infrastructure of the power network to reduce the physical damage from the disaster.

While hardening measures may be able to provide more resiliency sometimes, these strategies are often costly and inefficient \cite{panteli2015grid}. On the other hand, operational preventive solutions aim to mitigate the impact of catastrophic events by optimizing resources and updating control actions with the latest available information \cite{7922545}. 
Since operational preventive strategies are cost-efficient and more flexible to implement, a wide range of research has been conducted on this topic. Over an extended period, preventive control actions were determined through a deterministic optimization problem where no uncertain events were considered. However, to achieve an improved resilient power system, it is essential to consider the future uncertain events in the problem formulation. The authors in \cite{Yuan_robust} formulate a two-stage robust optimization method to enhance resiliency where a multi-stage and multi-zone based uncertainty of natural disaster is designed. In \cite{ma_resilient}, a resilient system is developed as a two-stage stochastic mixed-integer problem against high impact low probability weather events. The study in \cite{ball2006rebuilding} presented a case on hurricane preparedness and the rebuilding of the electrical infrastructure along the Gulf Coast following Hurricane Katrina. \cite{yamangil2015resilient} formulates the optimal design of electrical distribution grids as a two-stage stochastic mixed-integer program, where potential damage from natural disasters is represented by a set of stochastic events. A sequential unit commitment strategy was deployed by \cite{wang2018resilience} to mitigate cascading failure during disruptions from extreme events. Authors in \cite{wu2020integrating} introduce a resilience-based microgrid design model that optimizes the location and sizing of distributed energy resources and reconfiguration techniques. In \cite{arab2015stochastic}, a stochastic resource allocation framework for system operators is established to minimize potential damage to power system components due to upcoming hurricanes while optimizing cost-effectiveness. A risk assessment approach to infrastructure technology planning to improve power supply resiliency to natural disasters and other critical events is discussed in \cite{kwasinski2010technology}. 

The papers discussed above emphasize adopting preventive measures against the uncertain behavior of natural disasters utilizing different optimization methods. However, focusing on a specific event or uncertainty often results in an overly conservative solution for preventive EMS. Models developed to address one type of uncertain event may prove inefficient when applied to other events. To address this problem, our previous work \cite{anam2023risk} presented a scenario-based optimization method where each component's probability of failure (PoF) was included in the problem formulation. Since the formulation doesn't rely on specific types of uncertain events, different extreme events like natural disasters, faults, or cyber-attacks can be incorporated into the formulation by mapping their impacts on the components. Nonetheless, the scenario-based optimization method used in the work may be inefficient for large-scale power systems due to a large number of scenario generations. In this paper, we enhance the previous formulation by incorporating conditional-value-at-risk (CVaR) \cite{uryasev2001conditional} \cite{rockafellar2000optimization}, enabling the efficient integration of uncertain scenarios into the problem formulation. 

The resilient solutions, e.g., preventive and restorative approaches, are primarily formulated and solved using diverse optimization solution methods. However, to incorporate uncertainty and nonlinearity for complex power systems, optimization methods often result in significantly high computational complexity and require scenario reduction algorithms \cite{zhou2021deep}. Moreover, these strategies typically rely on offline, fixed settings with explicit optimization, meaning they cannot adapt to online solutions or learn from new conditions. In recent years, different learning-based approaches like machine learning (ML) and reinforcement learning (RL) algorithms are gaining popularity over traditional optimization methods in achieving power system resilience \cite{xie2020review} \cite{kuznetsova2013reinforcement}.  

In \cite{li2021integrating}, the Q-learning algorithm is utilized to establish the component restoration order, and a linear optimization algorithm is applied to achieve the maximum power supply for the specified network structure. The Q-learning algorithm has also proven effective in tackling power system resilience challenges, including sequential restoration \cite{wu2019sequential} and analyzing grid vulnerability during extreme events \cite{paul2020identification}. A multi-agent reinforcement learning-based approach is proposed in \cite{linh_restoration} using a deep Q network (DQN) to determine the optimal sequence of the circuit breakers. Authors in \cite{rocchetta2019reinforcement} create a Q-learning-based RL framework for the optimal operation and maintenance management of power grids. However, since Q-learning is a value-based RL algorithm, it is not applicable for continuous and high dimensional power system-related problems \cite{zhou2021deep}. The authors introduce an Advantage Actor-Critic (A2C)-based DRL framework in \cite{dehghani2021intelligent} to enhance the long-term resilience of distribution systems through hardening strategies using a combination of a novel ranking method, neural networks, and reinforcement learning. Although the authors focus on optimal planning problems to enhance resiliency, these types of problems can be solved using traditional optimization methods with superior performance. In \cite{tightiz2021resilience}, deep deterministic policy gradient (DDPG) and soft actor-critic (SAC) methods were utilized to solve the high-dimensional and stochastic problem of the microgrid’s EMS to maximize the stakeholders' profit. However, the authors only considered renewable generation as the uncertain parameter in the formulation and didn't include any probabilities of catastrophic events. 

The RL application to improve power system resilience is still in its early stages of development. This work aims to address the limitations of the current work in this field by developing a resilient EMS based on the RL algorithm. In this paper, a preventive EMS framework is proposed where the PoF of each system component is considered to generate all possible disruption scenarios with their corresponding probability. To mitigate the computational challenges of handling excessive scenarios, a threshold probability value is applied. The objective function is then reformulated as CVaR of the uncertain scenarios. As a result, the optimization problem is converted into a minimization problem resulting in a reduction of the computational burden. The formulated problem is solved with an RL agent using the Proximal Policy Optimization (PPO) \cite{schulman2017proximal} algorithm. A notional MVDC ship system and a modified IEEE 30-bus system are used to train the model. A Markov environment is created for the systems and the agent learns to generate the control decisions by sequentially training through the Markov Decision Process (MDP). Moreover, during training, a unique adaptive method with variable time steps is implemented to ensure that the process continues until it approaches a near-optimal solution. The results demonstrate that the proposed RL-based approach successfully generates control decisions for varying load profiles while ensuring all operational and network constraints are satisfied.

The main contributions of the paper are mentioned below:

\begin{itemize}
    \item In this paper, a novel preventive EMS framework is proposed that incorporates future uncertain events of the distribution system. Instead of focusing on a specific catastrophic event, the proposed model considers the PoF of each system component that can be utilized to generate a wide range of uncertain scenarios. The objective function of the proposed problem is further reformulated as the CVaR of the scenarios that significantly reduces the computational complexity.  

    \item To the best of our knowledge, this work represents the first occasion where the PPO algorithm is applied to solve a CVaR-based stochastic optimization problem in the power system domain. The learning capability of the RL agent enhances the resiliency of the power network against future uncertain events, offering a significant advantage over conventional solution methods. Additionally, the RL agent’s ability to make real-time decisions during actual disruptions allows it to respond quickly compared to the traditional optimization approaches.
\end{itemize}

The remainder of the paper is structured as follows: In Section \ref{faultcontainment:section2}
the optimization problem for the preventive EMS is developed. Initially, a deterministic problem is formulated, and then the CVaR-based minimization problem is introduced by considering different uncertain events. In Section \ref{faultcontainment:section3}, the DRL-based solution technique is proposed to solve the formulated problem. The experimental results are demonstrated in Section \ref{faultcontainment:section4} for a notional MVDC ship system and IEEE 30-bus bus system. Finally, Section \ref{faultcontainment:section5} summarizes the paper’s achievements and future work.

\section{Proposed Formulation for Preventive EMS} \label{faultcontainment:section2}
In this section, the optimization model is formulated in two stages; in the first stage, a deterministic model for the EMS is developed; in the following stage, the CVaR-based minimization problem is introduced to consider the uncertain scenarios based on the PoF of the system component. 
\subsection{Deterministic Optimization Model} \label{faultcontainment:section2.1}
The objective of the proposed model is to maximize the load served according to their priority with the available DERs:
\begin{equation}
	\underset{P}{\max} \sum_{t=1}^{T} \sum_{i=1}^{B} \left(w_{i} P_{i}^t \right) \Delta t 
	\label{eq:faultcontainment:Eq1}
\end{equation}
where, P is a vector consisting of all loads of the distribution system; $P_{i}^t$ is the load served at the $i$-th bus at time t; $w_i$ is the constant weight of the load at the $i$-th bus; and B, T, and $\Delta t$ are the total number of buses, the entire operational planning horizon, and the duration of each time step, respectively.
In this study, the loads are classified into critical, semi-critical, and noncritical loads; however, load classifications can be adjusted according to the system requirements. The EMS needs to serve the critical loads first, then the semi-critical, and at last, the non-critical loads. The objective function can be written as follows: 
\begin{equation}
	\underset{P}{\max} \sum_{t=0}^{T} \sum_{i=1}^{B} \left( K_{c}P_{i,c}^t + K_{sc}P_{i,sc}^t + K_{nc}P_{i,nc}^t \right) \Delta t 
	\label{eq:faultcontainment:Eq1.1}
\end{equation}
where, $K_c, K_{sc}, K_{nc}$ are constant weights of critical loads, semi-critical loads, and non-critical loads, respectively $(K_c\gg K_{sc} \gg K_{nc})$. 

The following active and reactive power balance constraints are associated with the system: 
\begin{equation}
    P_{i,inj}^t = P_{i,g}^t - P_{i,l}^t
    \label{eq:faultcontainment:Eq2}
\end{equation}
\begin{equation}
    Q_{i,inj}^t = Q_{i,g}^t - Q_{i,l}^t
    \label{eq:faultcontainment:Eq3}
\end{equation}
Depending on the nature of the distribution system, the AC or DC power flow equations should be used as constraints. For the MVDC system, 
\begin{equation}
    P_{i,inj} = \sum_{k = 1}^B V_i V_k G_{ik}
    \label{eq:faultcontainment:Eq4}
\end{equation}
The reactive power is not considered for the MVDC system. For the MVAC system:
\begin{equation}
    P_{i,inj} = \sum_{k = 1}^B V_i V_k (G_{ik} cos\theta_{ik} + B_{ik} sin\theta_{ik})
    \label{eq:faultcontainment:Eq4.1}
\end{equation}

\begin{equation}
    Q_{i,inj} = \sum_{k = 1}^B V_i V_k (G_{ik} sin\theta_{ik} + B_{ik} cos\theta_{ik})
    \label{eq:faultcontainment:Eq5}
\end{equation}

The following constraints should be included in the problem formulation to maintain the operational limits of the system: 

\begin{equation}
    P_{i,g}^{min} \leq P_{i,g}^t \leq P_{i,g}^{max},
    \label{eq:faultcontainment:Eq6}
\end{equation}

\begin{equation}
    Q_{i,g}^{min} \leq Q_{i,g}^t \leq Q_{i,g}^{max},
    \label{eq:faultcontainment:Eq7}
\end{equation}

\begin{equation}
    V_{i}^{min} \leq V_{i}^t \leq V_{i}^{max},
    \label{eq:faultcontainment:Eq8}
\end{equation}

\begin{equation}
    \theta_{i}^{min} \leq \theta_{i}^t \leq \theta_{i}^{max},
    \label{eq:faultcontainment:Eq9}
\end{equation}

\begin{equation}
    -P_{ik}^{lim} \leq P_{ik}^t \leq P_{ik}^{lim},
    \label{eq:faultcontainment:Eq10}
\end{equation}

\begin{equation}
    -Q_{ik}^{lim} \leq Q_{ik}^t \leq Q_{ik}^{lim},
    \label{eq:faultcontainment:Eq10.1}
\end{equation}
where (\ref{eq:faultcontainment:Eq6}) and (\ref{eq:faultcontainment:Eq7}) represent the generators' real and reactive power generation limits, (\ref{eq:faultcontainment:Eq8}) and (\ref{eq:faultcontainment:Eq9}) are the voltage and voltage angle limits, and (\ref{eq:faultcontainment:Eq10}) and (\ref{eq:faultcontainment:Eq10.1}) are the line limits for real and reactive power. 

The following ESS constraints are also included in the optimization problem:
\begin{equation}
    E_{i,b}^{t} = E_{i,b}^{t-1} - \eta_b P_{i,b}^{r,t} \triangle t
    \label{eq:faultcontainment:Eq12}
\end{equation}

\begin{equation}
    -P_{i}^{C,max} \leq P_{i,b}^{r,t} \leq P_{i}^{D,max},
    \label{eq:faultcontainment:Eq13}
\end{equation}

\begin{equation}
    \sum_{t=0}^{T} P_{i,b}^{r,t} = 0,
    \label{eq:faultcontainment:Eq14}
\end{equation}

\begin{equation}
    SOC_{i}^{min} \leq SOC_{i}^{t} \leq SOC_{i}^{max},
    \label{eq:faultcontainment:Eq15}
\end{equation}
where (\ref{eq:faultcontainment:Eq12}) indicates the energy conservation constraint, (\ref{eq:faultcontainment:Eq13}) is the limit for charging or discharging rate, and (\ref{eq:faultcontainment:Eq15}) is the state of charge (SOC) limit of the ESS. For the ESS, although the SOC can vary from 0 to 1 (0\% to 100\%), fully discharging can damage the battery permanently and shorten the life cycle of the battery \cite{noauthor_battery_nodate}. In this paper, the minimum SOC is selected as 0.2 (20\%). Eq.(\ref{eq:faultcontainment:Eq14}) ensures that the sum of the total charging and discharging power over a planning period will be zero, which helps the system to recharge the battery before the next planning cycle.

In addition, to enhance the resiliency of the system, the power generation is shifted towards more robust sources. The robustness of a power generator or converter can be expressed by a set of constants, $K^g = \{k_1^g, k_2^g, ...,k_n^g\}$, which are inversely proportional to the PoFs of the sources. To incorporate this into the formulation, the following constraint is included: 

\begin{equation} 
    k_1^g P_{1,g}^t = k_2^g P_{2,g}^t = ..... = k_N^g P_{N,g}^t
    \label{eq:faultcontainment:Eq00}
\end{equation}
Eq. (\ref{eq:faultcontainment:Eq00}) implies that power sources with higher $k_N^g$ values (corresponding to lower PoFs) contribute more to the total generation compared to the lower $k_N^g$ values (or high PoFs). As a result, the system will be less dependent on the vulnerable generators with high PoFs. 

\subsection{Reformulation using Conditional-Value-at-Risk (CVaR)} \label{faultcontainment:section2.2}

In this subsection, the optimization model is reformulated from a maximization problem to a CVaR-based minimization problem. CVaR specifically addresses high-impact-low probability events and minimizes the loss associated with these extreme scenarios. It is defined in relation to VaR (Value at Risk). By definition, for a given probability level $\alpha$, the $\alpha$-VaR of a portfolio represents the minimum value $\beta$ such that the portfolio's loss will not exceed $\beta$ with a probability of $\alpha$. In contrast, the $\alpha$-CVaR is the expected loss conditional on the losses exceeding $\beta$ \cite{rockafellar2000optimization}.  

Conditional Value at Risk (CVaR) at a confidence level \(\alpha\), measuring the risk of extreme losses, is defined as:

\[
\text{CVaR}_{\alpha}(X) = \mathbb{E}\left[ X \mid X \geq \text{VaR}_{\alpha}(X) \right]
\]
where, \(X\) is the random variable representing the loss, \(\text{VaR}_{\alpha}(X)\) is the Value-at-Risk at confidence level \(\alpha\), defined as:
\[
\text{VaR}_{\alpha}(X) = \inf \left\{ x \in \mathbb{R} \mid F_X(x) \geq \alpha \right\}
\]
where \(F_X(x)\) is the cumulative distribution function of \(X\). \(\text{VaR}_{\alpha}(X)\) is the value below which \(\alpha\) fraction of the losses fall. \(\text{CVaR}_{\alpha}(X)\) is the expected loss given that the loss exceeds \(\text{VaR}_{\alpha}(X)\). It provides a measure of the tail risk and is particularly useful for assessing the risk of extreme losses.

The optimization model in Section \ref{faultcontainment:section2.2} can be reformulated with the CVaR by introducing the loss function in terms of load variables. For a finite set of uncertain scenarios, $S^u = \{s_1^u, s_2^u, s_3^u,...s_n^u\}$, the loss function can be defined as: 

\begin{equation}
    P_{loss}(s_i^u) = E[P_{load}(s_i^u)] - P_{load}(s_i^u)
    \label{eq:faultcontainment:Eq17}
\end{equation}
where $E[P_{load}(s_i^u)]$ is the expected load served and $P_{load}(s_i^u)$ is the actual load served in the $i$-th scenario, where, 
\begin{multline}
    P_{load}(s_i^u) = \sum_{t=0}^{T} \sum_{i=1}^{B}( K_{c}P_{i,c}^t (s_i) + K_{sc}P_{i,sc}^t (s_i) \\ + K_{nc}P_{i,nc}^t(s_i) \Delta t  
\end{multline}
According to the definition, the CVaR can be expressed as: 

\begin{equation}
    CVaR = \frac{1}{1 - \alpha} \sum_{s_i^u \in S^u} P_{loss}(s_i^u) p(s_i) |{P_{loss}(s_i^u) \geq VaR}
    \label{eq:faultcontainment:Eq18}
\end{equation}
where $p(s_i)$ is the probability of $i$-th scenario. 

So, in the final problem formulation, (\ref{eq:faultcontainment:Eq1}) will be replaced with (\ref{eq:faultcontainment:Eq18}) and the maximization problem will be converted into a minimization problem: 
\begin{center}
   $min$ (\ref{eq:faultcontainment:Eq18})
   
   s.t. (\ref{eq:faultcontainment:Eq2}), (\ref{eq:faultcontainment:Eq3}), (\ref{eq:faultcontainment:Eq6}) - (\ref{eq:faultcontainment:Eq00}). 
\end{center}

\section{DRL Framework for Preventive EMS} \label{faultcontainment:section3}

\subsection{Overview of Deep Reinforcement Learning(DRL)}
In this literature, a DRL-based method is implemented to solve the optimization problem formulated in Section \ref{faultcontainment:section2}. An RL algorithm consists of two main components: an environment and one or more agents. The agent continuously interacts with the environment and adjusts its policy to take optimal control actions \cite{sutton2018reinforcement}.

Sequential interactions between the agent and the environment can be represented by a Markov Decision Process (MDP). MDP can be expressed as follows: 

\begin{center}
    $MDP = f(S, A, P, R, \gamma)$
\end{center}

where, $S = \{s_1, s_2, ...,s_n\}$ is the finite set of states, $A = \{a_1(S), a_2(S),..., a_n (S)\}$ represents the set of actions taken by the agent in the states of $S$, $P$ is the set of transition probability $p(s, a, s')$ of taking action $a$ to transit from state $s$ to $s'$, $R$ represents the reward function, and $\gamma$ is the discount factor. 

During the training process, the agent generates a set of actions provided to the environment to produce the corresponding observations and rewards. The agent's goal is to optimize the policy so that the cumulative discounted reward $R$ is maximized for a set of actions and observations. The cumulative discounted reward $R$ can be expressed as follows: 

$R_t = r(s_t, a_t) + \gamma r(s_{t+1}, a_{t+1}) + \gamma^2 r(s_{t+2}, a_{t+2}), ... = \sum_{t'=0}^{T-1} \gamma^{n-1} r(s_{t+t'}, a_{t+t'})$

The optimal algorithm of the agent maps a set of states to a set of actions by maximizing the reward. RL algorithm primarily involves two approaches; value-based methods and policy-based methods \cite{nachum2017bridging}. In the value-based method, the agent learns a value function that represents the state-action values, and based on the values, the agent chooses the best action. On the other hand, in the policy-based method, the agent directly learns the policy that maps the states into actions. In power system control, the action and observation spaces are typically continuous and high-dimensional, making value-based algorithms impractical to use \cite{zhou2021deep}. As a result, the policy-based DRL algorithm is used to solve these types of problems. 

In this paper, a Proximal Policy Optimization(PPO) algorithm is used to train the agent. 

\begin{figure*}[htb]
\centering
\includegraphics[width=\textwidth]{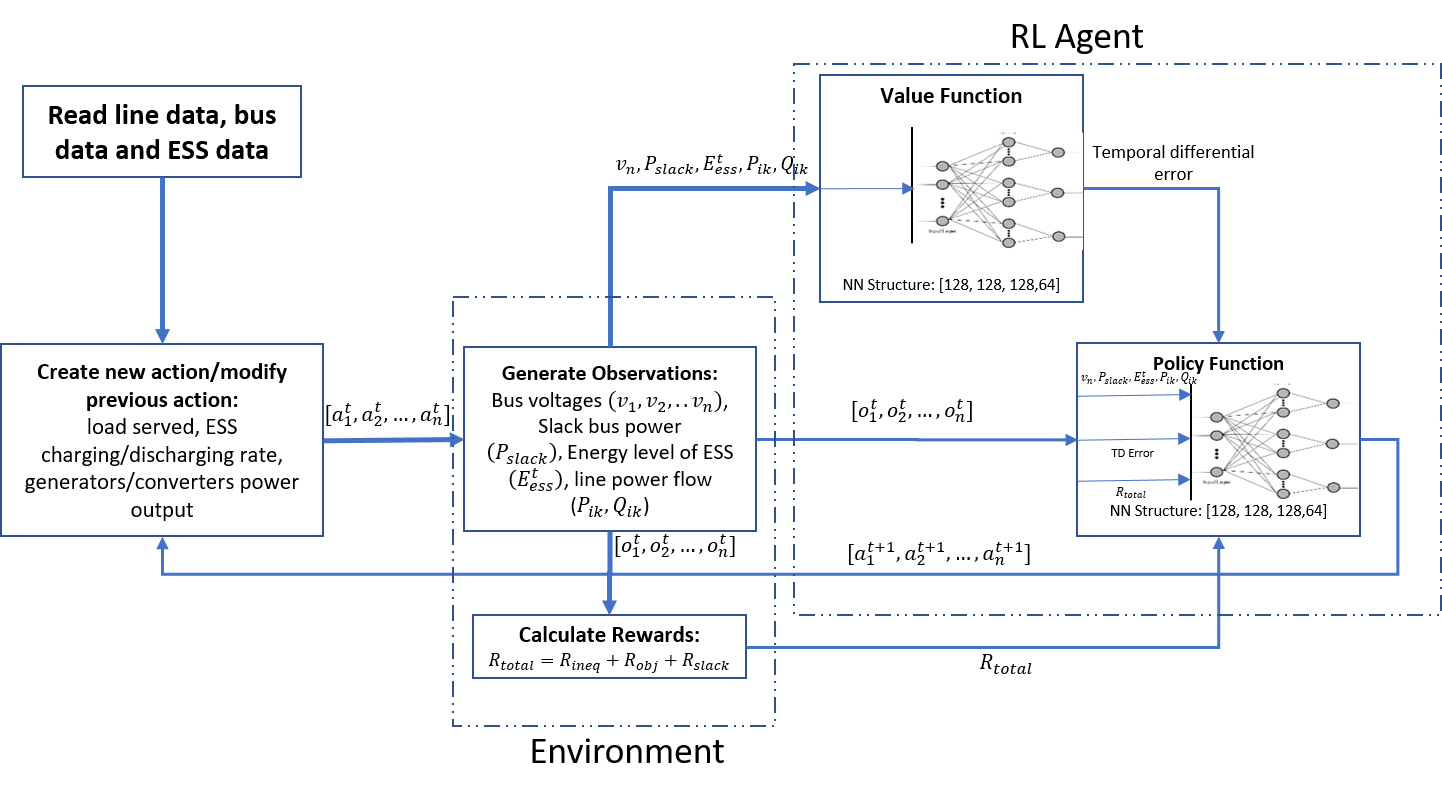}
\caption{Proposed DRL-based framework with PPO algorithm for the Preventive EMS}
\label{fig:solution}
\end{figure*}

\subsection{Proximal Policy Optimization (PPO) Algorithm}

PPO is an advanced reinforcement learning algorithm designed to enhance policy gradient methods. It seeks to balance exploration and exploitation while avoiding significant policy shifts by restricting the update magnitude. PPO optimizes the expected reward using a clipped surrogate objective using stochastic gradient ascent \cite{schulman2017proximal}. 

The PPO algorithm involves the following components:

\begin{itemize}
    \item \textbf{Policy}: \( \pi_\theta(a|s) \) represents the policy that outputs the probability of taking action \( a \) in state \( s \) with parameters \( \theta \).
    \item \textbf{Value function}: \( V_\phi(s) \) estimates the expected return from state \( s \) with parameters \( \phi \).
    \item \textbf{Advantage function}: The advantage function measures how much better an action is compared to others in a given state.
    \[
    A_t = R_t + \gamma V_\phi(s_{t+1}) - V_\phi(s_t)
    \]
    where \( R_t \) is the reward, \( \gamma \) is the discount factor, and \( V_\phi(s) \) is the value function.
\end{itemize}

The objective function of PPO aims to optimize the policy using the ratio between the new and old policy probabilities:

\[
r_t(\theta) = \frac{\pi_\theta(a_t|s_t)}{\pi_{\theta_{\text{old}}}(a_t|s_t)}
\]

The PPO loss function is defined as:

\[
L^{\text{PPO}}(\theta) = \mathbb{E}_t \left[ \min \left( r_t(\theta) A_t, \, \text{clip}(r_t(\theta), 1-\epsilon, 1+\epsilon) A_t \right) \right]
\]

Here, \( \epsilon \) is a small constant that controls how much the policy is allowed to change at each update. The clipping ensures that the updates to the policy do not deviate too much, thus preventing large policy updates that could destabilize learning.

The loss for the value function is the squared error between the predicted value and the actual return:

\[
L_{\text{VF}}(\phi) = \frac{1}{2} \mathbb{E}_t \left[ \left( V_\phi(s_t) - R_t \right)^2 \right]
\]

To encourage exploration, PPO adds an entropy term to the objective function. This term increases the policy's randomness during training:

\[
L_{\text{entropy}} = \mathbb{E}_t \left[ -\pi_\theta(a_t|s_t) \log \pi_\theta(a_t|s_t) \right]
\]

The total loss for PPO is a combination of the policy loss, value function loss, and entropy bonus:

\[
L_{\text{total}} = L^{\text{PPO}}(\theta) + c_1 L_{\text{VF}}(\phi) - c_2 L_{\text{entropy}}
\]

where \( c_1 \) and \( c_2 \) are coefficients that balance the contributions of the value function loss and entropy bonus, respectively.

\subsection{Proposed Control Algorithm for Preventive EMS}

To solve the problem using the RL algorithm, the formulated optimization problem should be redefined as an MDP environment with corresponding action space, observation space, and reward function. In the proposed method of this paper, the RL components are modeled as follows: 

\begin{itemize}
    \item Environment: The distribution system is the environment for the formulated problem where the agent interacts to generate the control actions. The distribution network includes distributed generators with multiple power converters, ESS, loads, and lines. In power system, it is very crucial to maintain the network and operational constraints while running the EMS. Violation in one or more constraints can lead to system stability or even failure. There are two types of constraints in the problem formulated in Section \ref{faultcontainment:section2}: (\ref{eq:faultcontainment:Eq2}), (\ref{eq:faultcontainment:Eq3}), (\ref{eq:faultcontainment:Eq12}), and (\ref{eq:faultcontainment:Eq14}) are the equality constraints; and (\ref{eq:faultcontainment:Eq6}) - (\ref{eq:faultcontainment:Eq10.1}), (\ref{eq:faultcontainment:Eq13}) and (\ref{eq:faultcontainment:Eq15}) are the inequality constraints. For the inequality constraints, certain penalties (or negative rewards) are assigned for the violations so that the agent gradually learns to maintain the constraints. On the other hand, maintaining the equality constraints with the RL agent is more complicated. Since, for all learning-based methods, there is a certain error margin, only introducing penalties will not ensure the maintenance of these constraints. In this paper, we adopted two separate methods to maintain the power flow and ESS equality constraints, respectively. 

    A power flow function is presented where the control values (actions) are provided to solve the power flow equations. A slack bus is introduced inside the function that can adjust power to maintain the equality constraints. The slack bus power should be within a specific range depending on the rated net power input of that bus: 
    
    \begin{center}
        $P_{slack}^{min} \leq P_{slack} \leq P_{slack}^{max}$
    \end{center}

    The ESSs are controlled in three steps: first, the value of the charging/discharging power of each ESS is provided by the agent; second, the remaining stored energy of the ESS is updated according to (\ref{eq:faultcontainment:Eq12}); and third, the next ESS dispatch limit is determined by the following equation: 

    \begin{equation}
        \begin{aligned}
            min(P_i^{C,max} \triangle t, E_{i,b}^t-E_{i,b}^{min}) \leq P_{i,b}^{r,t+1} \\ \leq min(P_i^{D,max} \triangle t, E_{i,b}^{max} - E_{i,b}^t)
        \label{eq:faultcontainment:Eq21}
        \end{aligned}
    \end{equation}

    Eq. (\ref{eq:faultcontainment:Eq21}) ensures that the dispatch for the next time step will not exceed the remaining energy level of the ESS. 
    
    \item Action space: A set of actions, $A = [a_1^t, a_2^t, ..., a_n^t]$, is generated by the agent based on exploration or exploitation. Three types of actions are generated in the proposed framework: 
    \begin{enumerate}
        \item The amount of load served: $p_1, p_2, p_3,....$
        \item ESS charging/discharging power: $P_{i,b}^{r,t}$
        \item Output power from the generators/converters: $P_{i,g}^t$
    \end{enumerate}
    
    \item Observation space: When the actions are supplied to the environment, the required observation set, $O = [o_1^t, o_2^t,..., o_n^t]$, are generated. The observations considered in this paper are: bus voltage $v_i$ and voltage angle $\theta_i$, power at slack bus $P_{slack}^t$, real and reactive power flowing through each line $P_{ik}^t$ and $Q_{ik}^t$, respectively, and energy level of each ESS $E_{i,b}^t$. 
    \item Reward function: The accuracy of the RL algorithm largely depends on a well-defined reward function. The reward function developed in this literature has three parts: 
    \begin{enumerate}
        \item Positive reward for load-served, $R_{obj}$: This reward is the value of the objective function. The objective of the EMS is to serve as much loads as possible according to their weight. The more loads the agent can supply, the higher the value of $R_{obj}$ it will receive. 
        \item Negative rewards for inequality constraints violation, $R_{ineq}$: If the agent fails to maintain the inequality constraints, it will receive a negative reward (or penalty) for each violation. $R_{ineq}$ can be defined as follows: 

        $R_{ineq} = R_{v|\theta} + R_{P|Q}$
        where $R_{v|\theta}$ is the penalty for voltage or voltage angle violations of the buses, and $R_{P|Q}$ is the penalty for real or reactive power limit violation of the lines. 
        \item Rewards for slack bus power, $R_{slack}$: Since the slack bus power should be maintained between $P_{slack}^{min}$ and $P_{slack}^{max}$, $R_{slack}$ is defined as follows: 
        \[
            R_{slack} = 
            \begin{cases} 
                k_1 \times (P_{slack} - P_{slack}^{min}) & \text{if } P_{slack} \leq P_{slack}^{min}, \\
                k_2 \times (P_{slack}^{max} - P_{slack}) & \text{if } P_{slack} \geq P_{slack}^{max}, \\
                0 & \text{Otherwise }
            \end{cases}
        \]
        where $k_1$ and $k_2$ are positive constants. The violation of the slack bus power limit will result in a penalty for the agent and higher violation will cause higher penalty. 
    \end{enumerate}
    The total reward will be: 

    $R_{total} = R_{obj}+R_{ineq}+R_{slack}.$
\end{itemize}

\section{Case Studies} \label{faultcontainment:section4}

\begin{figure}[t]
\centering
\includegraphics[scale=0.11]{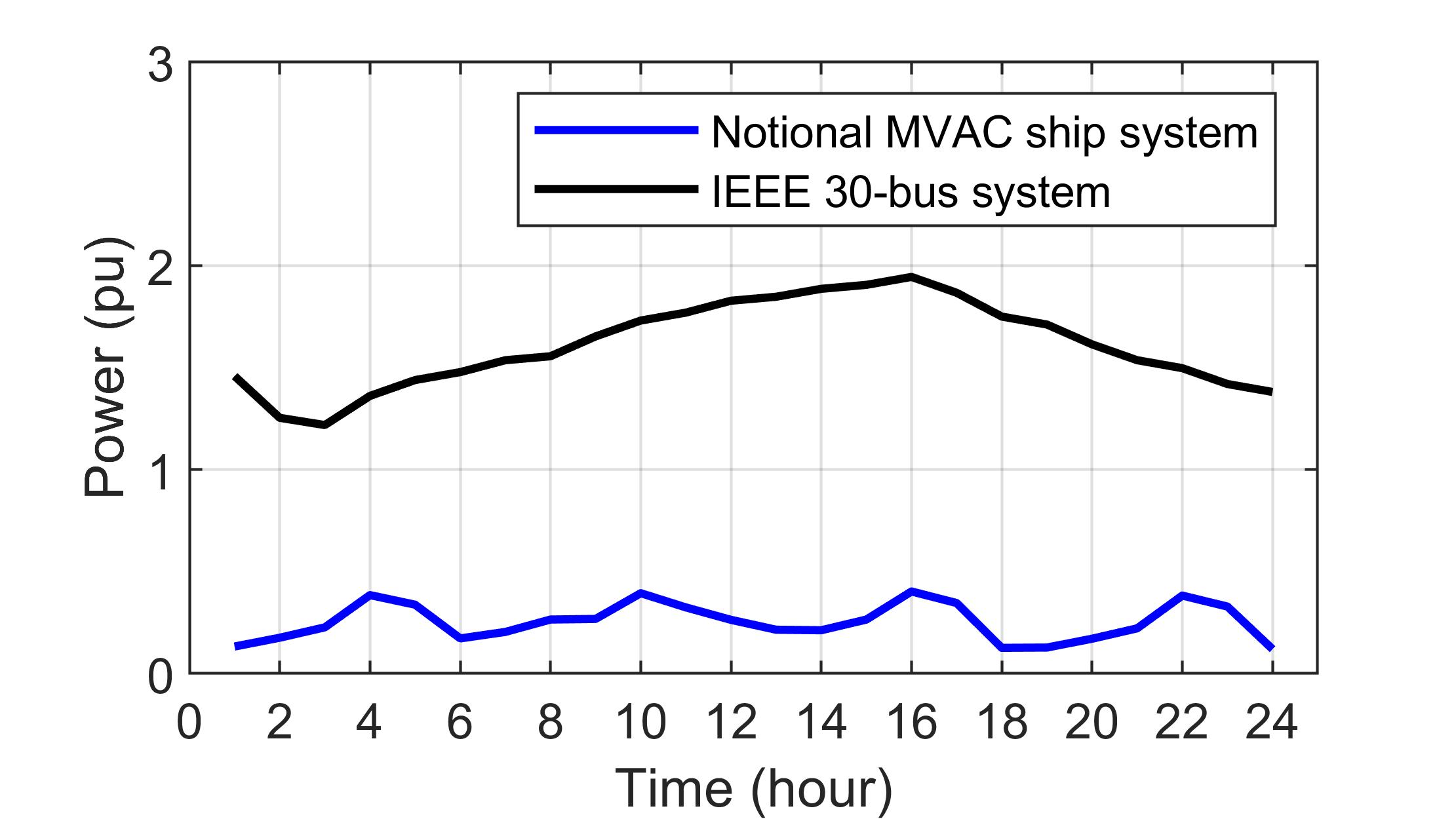}
\caption{24-hour load profile of notional ship system and IEEE 30-bus system}
\label{fig:load_profile}
\end{figure}

The proposed Reinforcement Learning (RL)-based approach is validated using a notional 12-bus MVDC ship system and a modified IEEE 30-bus system. The load profiles for both systems are depicted in Fig. \ref{fig:load_profile}. All training and testing processes are conducted on a system equipped with an Intel(R) Core(TM) i7-10700 CPU (2.90 GHz) and 16.0 GB of RAM. Then the obtained results are benchmarked against a conventional scenario-based optimization method. 

\subsection{Scenario Generation Based on PoFs} \label{faultcontainment:section4.0}

A set of scenarios is generated based on the PoFs of the system component. For a system with $n$ components, the total number of possible scenarios is given by $2^n$, as each component can either succeed or fail, leading to a binary combination of states. However, not all scenarios are practically significant, as some have extremely low probabilities, which may result in negligible impact to the EMS.

To refine the analysis and focus on more meaningful scenarios, a probability threshold is applied. This threshold eliminates scenarios with the extreme low probabilities to keep the solution more practical. By filtering out these low-value scenarios, we concentrate on the more impactful events, ensuring that the analysis is both computationally efficient and prioritizing scenarios that are most relevant to the system's performance or risk assessment.

\begin{table}[b]
	\renewcommand{\arraystretch}{1.3}
	\caption{Converter Data for Notional MVDC Ship System}
	\label{tab:MVDC_converter_data}
	\centering
	\begin{tabular}
		{
			>{\centering\arraybackslash}m{0.08\textwidth}
			>{\centering\arraybackslash}m{0.08\textwidth}
			>{\centering\arraybackslash}m{0.14\textwidth}
			>{\centering\arraybackslash}m{0.08\textwidth}
		}
		\hline
		Types 
		& Number of Converters
		& Capacity (MW)
		& PoF (from converter 1 to n) 
		\\
		\hline
		ATG-1 & 2 & 2.6 & .05, .075 \\
        ATG-2 & 2 & 2.6 & .05, .05 \\
        MTG-1 & 5 & 8.2 & .05, .05, .05, .075, .05 \\
        MTG-2 & 5 & 8.2 & .025, .05, .05, .05, .05 \\
		\hline\hline
        \end{tabular}
\end{table}

\begin{table}[b]
	\renewcommand{\arraystretch}{1.3}
	\caption{ESS data for notional MVDC ship system and IEEE 30-bus system}
	\label{tab:ESS_data}
	\centering
	\begin{tabular}
		{
			>{\centering\arraybackslash}m{0.1\textwidth}
			>{\centering\arraybackslash}m{0.05\textwidth}
			>{\centering\arraybackslash}m{0.05\textwidth}
			>{\centering\arraybackslash}m{0.1\textwidth}
			>{\centering\arraybackslash}m{0.08\textwidth}
		}
		\hline
		System 
		& Number of ESS
		& Capacity (MWh) 
		& Maximum Charging/discharging rate (MW/h) 
		& Miniumum SOC (\%) \\
		\hline
		MVAC ship system & 8 & 2.2 & 10 & 20\\
        IEEE 30-bus system & 6 & 15 & 5 & 20\\
		\hline\hline
        \end{tabular}
\end{table}

\subsection{Notional 12-Bus MVDC Ship System} \label{faultcontainment:section4.1}

\begin{figure}[t]
\centering
\includegraphics[scale=.26]{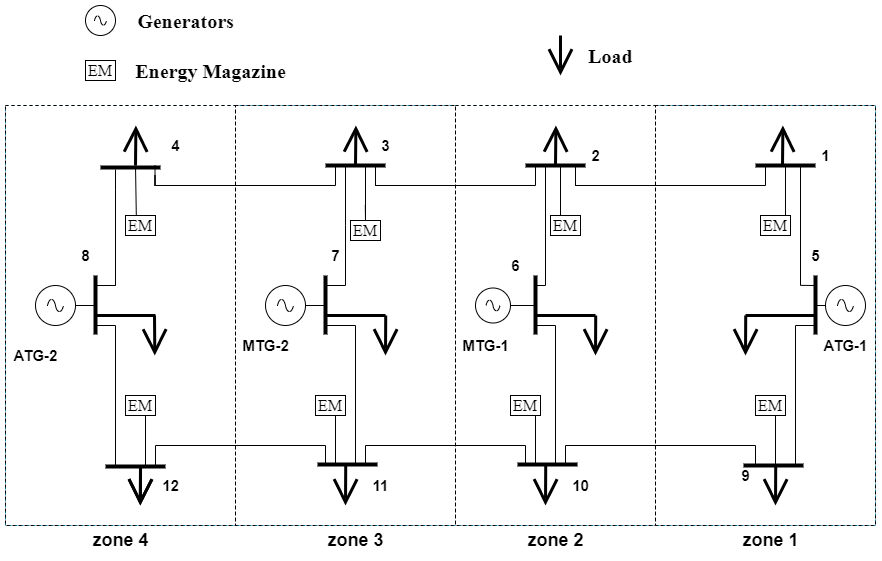}
\caption{Line diagram of the notional MVDC ship system model}
\label{fig:MVAC_model}
\end{figure}

The notional four-zone 12 bus MVDC ship system \cite{MVDC_ship} (shown in Fig. \ref{fig:MVAC_model}) has two main gas turbine generators (MTG) and two auxiliary gas turbine generators (ATG). TABLE \ref{tab:MVDC_converter_data} represents the generator parameters of the system. Additionally, the system includes 8 ESS (with each zone containing 2 ESS) and multiple loads, including 2 propulsion motor modules (PMM) at buses 6 and 7 and AC load centers (ACLC) distributed across buses 1,2,3,4,9,10,11, and 12. The ESS data are provided in TABLE \ref{tab:ESS_data}. 

\begin{figure}[t]
\centering
\includegraphics[scale=1]{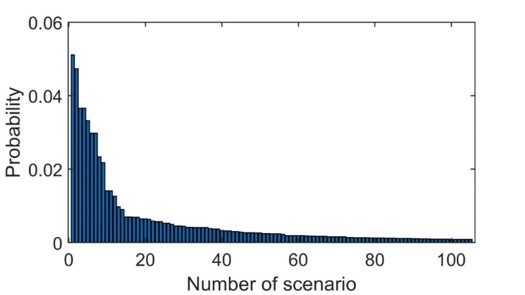}
\caption{Number of scenarios after applying threshold}
\label{fig:MVDC_scenario}
\end{figure}

The RL agent based on the PPO algorithm is trained on the MVDC ship system for 450,000 episodes. Determining the appropriate number of time steps in each episode is a challenging task for the learning-based methods. The agent often stops training because of a limited number of time steps while one or more constraints remain unsatisfied. On the other hand, using a high value of time steps can lead to extensive training time. To address this, a novel variable time step strategy is implemented in this work. In this method, the agent continues to train until all constraints are satisfied. As a result, during the training procedure, the number of time steps can be different for different episodes. The number of final scenarios after applying a threshold value of .0005 can be observed from Fig. \ref{fig:MVDC_scenario}. 

\begin{figure}[t]
\centering
\includegraphics[scale=0.40]{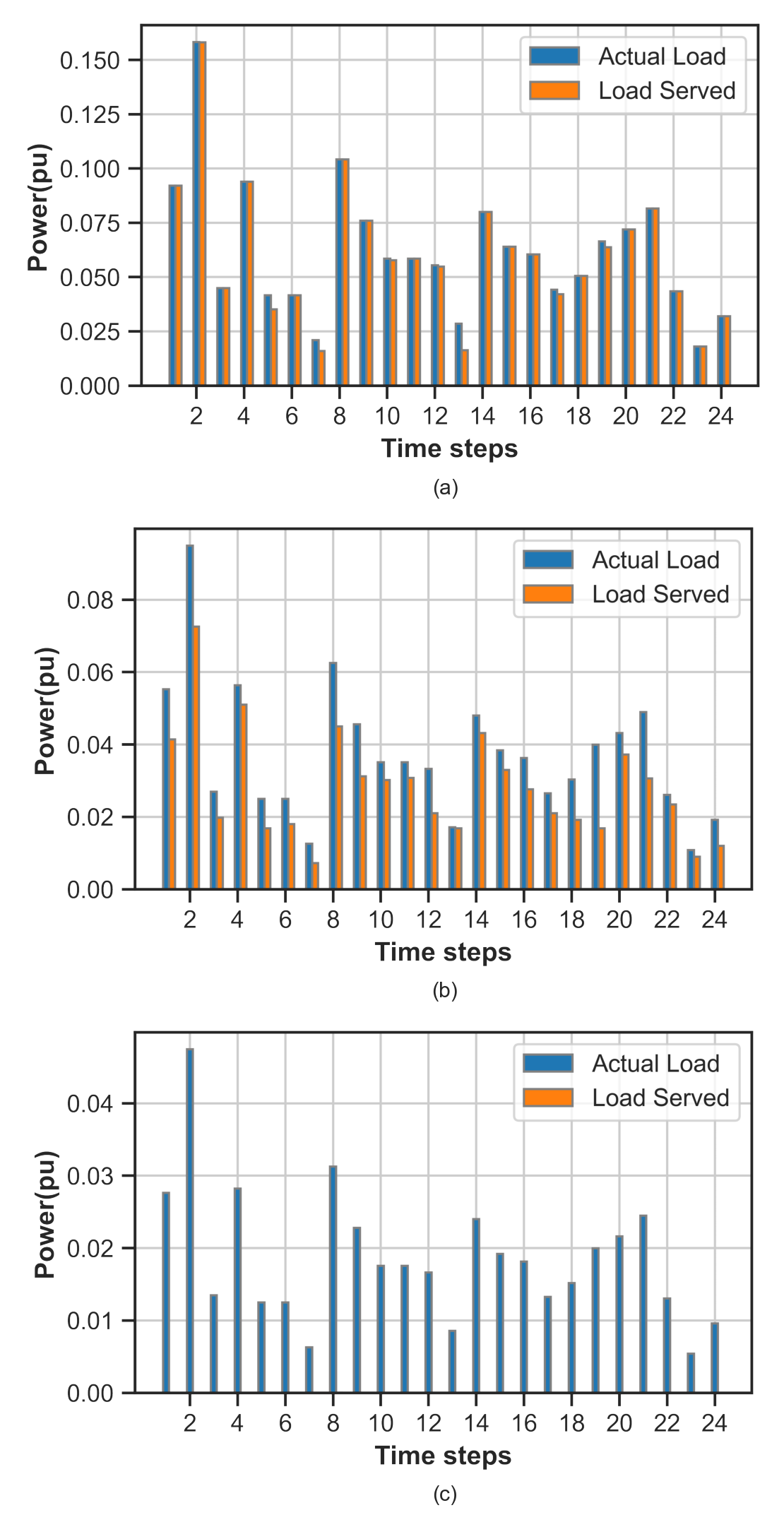}
\caption{Results for MVDC ship system (a) Critical load profile vs. critical load served, (b) Semi-critical load profile vs. semi-critical load served, (c) Noncritical load profile vs. non-critical load served. }
\label{fig:MVDC_load_served}
\end{figure}

\begin{figure}[t]
\centering
\includegraphics[scale=1]{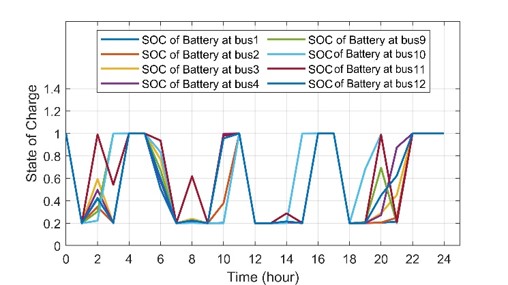}
\caption{SOC of the energy storage systems (ESS)}
\label{fig:MVDC_soc}
\end{figure}

\begin{figure}[t]
\centering
\includegraphics[scale=.15]{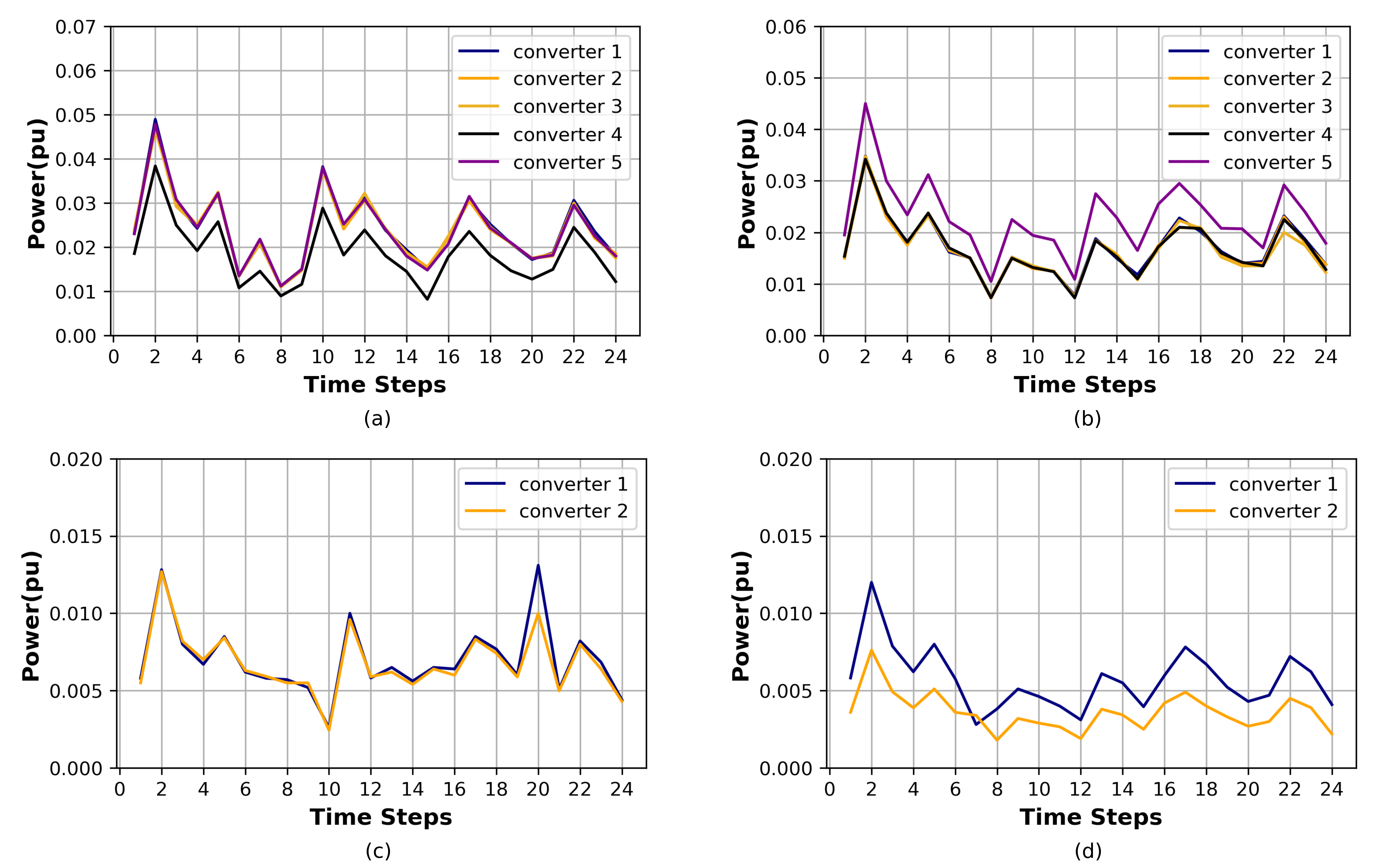}
\caption{Converter outputs of (a) MTG 1, (b) MTG 2, (c) ATG 1, (d) ATG 2}
\label{fig:MVDC_converter}
\end{figure}

The results for the MVDC ship system are demonstrated in Fig. \ref{fig:MVDC_load_served}. It can be observed that the agent prioritizes the critical loads to serve over the planning horizon. On the other hand, some of the semi-critical loads and all the non-critical loads have been curtailed. The SOC of the ESS can be observed from Figs. \ref{fig:MVDC_soc}. Additionally, the generation of the system is shifted towards the robust converter as depicted in Fig. \ref{fig:MVDC_converter}. This figure indicates that, converters with high PoFs contribute less towards the total generation compared to the converters with low PoFs. 

\subsection{IEEE 30-bus system} \label{faultcontainment:section4.2}

\begin{figure}[t]
\centering
\includegraphics[scale=0.4]{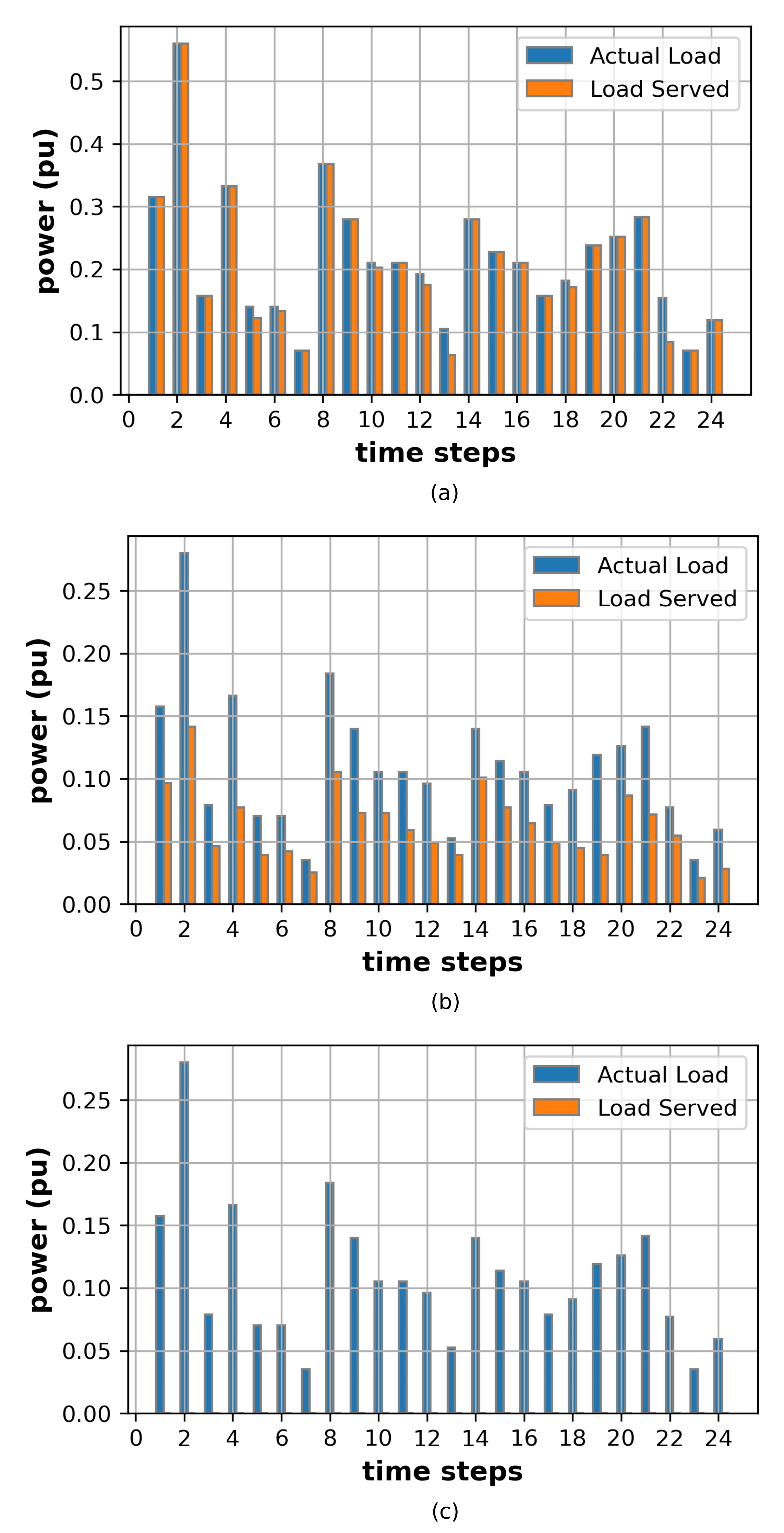}
\caption{Results for IEEE 30-bus system (a) Critical load profile vs. critical load
served, (b) Semi-critical load profile vs. semi-critical load served, (c) Noncritical load profile vs. non-critical load served. }
\label{fig:IEEE_30_load_served}
\end{figure}

The IEEE 30-bus system \cite{IEEE30bus} has 6 generators, 41 lins and 21 load buses. The system is further modified by adding six (6) ESSs, shown in \ref{tab:ESS_data}. The RL agent is trained on the IEEE 30-bus system for 600,000 episodes. Since the scenario generation and variable time step implementation techniques are similiar to those used for the MVDC ship system, their detailed procedures are not included in this section. 
The EMS optimized the system to supply the critical loads first, then the semi-critical loads, and at last, the non-critical loads. The results are shown in Fig. \ref{fig:IEEE_30_load_served}. The figure shows that the system supplied the majority of the critical loads, although the system should supply all the critical loads before supplying any semi-critical loads. The reason is the active power flow limit between two buses. When a line reaches the limit for active power flow, the system supplies power through other lines to a different bus, even if it has less important loads.

\subsection{Result Comparison}

\begin{figure}[t]
\centering
\includegraphics[scale=.7]{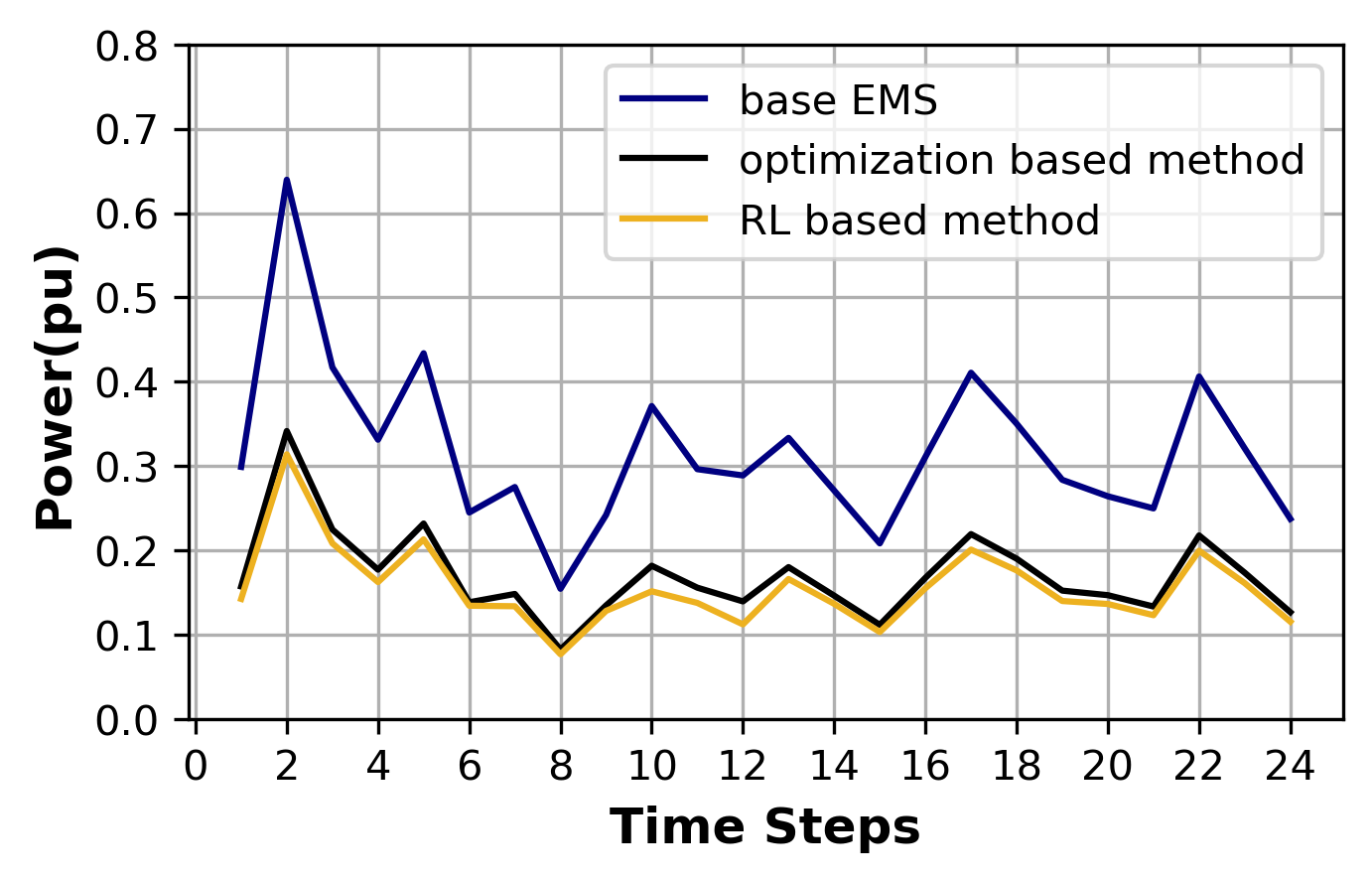}
\caption{Result comparison between base EMS, optimization-based method, and RL-based method}
\label{fig:comparisons}
\end{figure}

For validation, the results obtained by the RL algorithm are compared with a base EMS and a traditional scenario-based optimization model. The comparison can be observed from Fig. \ref{fig:comparisons}. 

The base EMS represents the state of the system without any resilience measures, where no uncertain scenarios are considered and the generators operate at their maximum capacity. Although the base EMS supplies a higher amount of load compared to the resilient EMS, this type of system condition can lead to cascading failure during the events of natural disasters or faults. The gap between the base EMS and the resilient EMS shown in Fig. \ref{fig:comparisons} depends on the system's vulnerability. A system with components of high PoFs is more vulnerable, leading to the resilient EMS serving a lower amount of load to enhance system resiliency. In contrast, improving the security or infrastructure of the power network will improve the robustness of the system, resulting in moving the curve closer to the base EMS.

The RL-based resilient EMS is also compared with a traditional scenario-based convex optimization method. For both optimization-based and RL-based methods, the same PoF values are used to obtain the results. It is clear that traditional optimization methods demonstrate slightly better performance in terms of maximizing the load served, ensuring that more demand is met in the system. However, in terms of computational time, RL algorithm outperforms the optimization method as demonstrated in TABLE \ref{tab:simulation_time}. It should be noted that, although the training time for the DRL-based methods can be high, once the agent is trained for a specific system, the testing time is considered as the simulation time. 

Conventional optimization methods often rely on well-defined mathematical models and are capable of finding globally optimal solutions when system conditions are static or well-understood. On the other hand, Reinforcement Learning (RL) methods offer significant advantages in terms of adaptability and computational efficiency. RL is highly effective in dynamic and uncertain environments where system conditions fluctuate over time. These characteristics make the DRL-based algorithm more suitable to apply in an uncertain environment. 

\begin{table}[b]
	\renewcommand{\arraystretch}{1.3}
	\caption{Simulation time comparison between DRL algorithm and traditional optimization method}
	\label{tab:simulation_time}
	\centering
	\begin{tabular}
		{
			>{\centering\arraybackslash}m{0.08\textwidth}
			>{\centering\arraybackslash}m{0.08\textwidth}
			>{\centering\arraybackslash}m{0.14\textwidth}
			>{\centering\arraybackslash}m{0.08\textwidth}
		}
		\hline
		System 
		& DRL Agent (s)
		& Optimization Method (s)
		\\
		\hline
		MVDC Ship System & 1.43 & 7.12\\
        IEEE 30-Bus System & 1.91 & 15.2 \\
		\hline\hline
        \end{tabular}
\end{table}

\section{Conclusion} \label{faultcontainment:section5}

The increasing complexity of the advanced power system network necessitates enhancing the resiliency of the system for a reliable and uninterrupted electricity supply. While developing the EMS with the goal of building a robust system, it is crucial to account for future uncertain events in the formulation. In this paper, a resilient preventive EMS framework is proposed, where the uncertain disruption scenarios are generated considering the PoFs of the system components. The optimization problem formulated as a CVaR minimization problem resulting in the reduction of the computational burden. The formulated problem is solved with a DRL-based method where the agent uses the PPO algorithm to generate the control decision. The DRL-based method offers more adaptability against the uncertain environment and faster decision making ability compared to the conventional method. Additionaly, the DRL-based framework can handle the nonlinear system directly the traditional optimization method requires complicated modification. 

Since the DRL-based approach for the power system-related problem is still an emerging sector, more research should be done to broaden the applications of this method. Although this literature develops a novel framework for a resilient EMS, there are still some limitations in terms of accuracy and scalability. Some other advanced methods like Multi-Agent RL (MARL) or Deep Graph RL (DGRL) can be explored to generate possible solution of these limitations.

\bibliographystyle{IEEEtran}
\bibliography{references}

\end{document}